\documentclass[11pt]{article}
\usepackage[dvips]{epsfig}

\begin{document}
\begin{center}
{\bf \Large PHOTOIONIZATION AND RECOMBINATION }\\
{\large Sultana N. Nahar\\Department of Astronomy, The Ohio State
University\\ Columbus, Ohio, USA 43210}
\end{center}
\begin{quotation}
\noindent {\bf Abstract}. Theoretically self-consistent calculations
for photoionization and (e~+~ion) recombination are described.
An identical eigenfunction expansion for the ion is employed in
coupled channel calculations for both processes, thus ensuring
consistently accurate cross sections and rates in an ab initio manner.
The theoretical treatment of (e~+~ion) recombination subsumes both
the non-resonant recombination ('radiative recombination'), and
the resonant recombination ('di-electronic recombination')
processes in a unified scheme. In addition to the total, unified
recombination rates, level-specific recombination rates and 
photoionization cross sections are obtained for a large number of 
atomic levels. Both relativistic Breit-Pauli, and non-relativistic 
LS coupling, calculations 
are carried out in the close coupling approximation using the R-matrix 
method. Although the calculations are computationally intensive, they
yield nearly all photoionization and recombination parameters needed for
astrophysical photoionization models with higher precision than hitherto
possible, estimated at about 10-20\% from comparison with
experimentally available data (including 'experimentally derived DR
rates'). Results are electronically available for over 40 atoms and
ions. Photoionization and recombination of He-, and Li-like C and
Fe are described for X-ray modeling. The unified method yields total and
complete (e+ion) recombination rate coeffcients, that can not otherwise be 
obtained theoretically or experimentally.

\end{quotation}


\section{Introduction}


 Although photoionization and recombination are inverse processes as
they occur in nature, they are usually treated in independent
theoretical frameworks. This basic inconsistency, directly related
to ionization balance in radiatively ionized media, and consequent
inaccuracies, propagate through to the photoionization models employed
in astrophysics. A further division, largely artificial, is made in
theoretical methods used to compute electron-ion recombination rates.
Two sets of data are usually calculated: (i) 'radiative recombination' 
(RR), calculated using background, or non-resonant, photoionization cross
sections, and (ii) 'di-electronic recombination' (DR) representing the
contribution of autoionizing resonances, first shown to be important by
Burgess (1964). That this procedure is not only theoretically
unsatisfactory, but also impractical in most cases, is seen from both 
theoretical calculations and experimental measurements of 
photoionization and recombination cross sections. The simple reason 
is that the resonances are inseparable from the background.
The cross sections contain, in general, extensive and
interacting Rydberg series of resonances; the non-resonant and resonant
contributions are not accurately separable (except, possibly, for 
few-electron, highly charged ions). The large number of
photoionization cross sections computed under the Opacity Project
exhibit these features in detail, particularly for many electron
systems ({\it The Opacity Project Team} 1995, 1996 - compiled 
publications and data) In addition, the cross sections for 
photoionization and recombination of excited states, particularly 
metastable levels, may contain even more complicated resonances than 
the ground state (Luo et al 1990). Experimentally, of course, the 
measurements {\it always} yield a combined '(RR + DR)' cross section 
(albeit in limited energy ranges usually accessible in
experimental devices).

Therefore a theoretical method that accounts for both the resonant and
the non-resonant recombination in a unified manner is desirable, and has
been developed (e.g. Nahar and Pradhan 1994, Zhang et al 1999),
based on the close coupling (CC) approximation using the
R-matrix method (Burke \& Seaton 1984, Berrington et al 1987, Hummer et al
1993) as used in the Opacity Project and the Iron Project (hereafter OP
and IP). Photoionization cross sections may be computed essentially 
for all bound states, level of excitation ($n,\ell,SL\pi,SLJ\pi)$, 
energy range, and with resolution of resonances. In principle, the 
cross section for the inverse photo-recombination process is given 
by detailed balance. However, since
recombination takes place to an infinite number of bound states of
(e~+~ion) system, it becomes impractical (and as it turns out,
unnecessary) to do so for the very higly excited levels above a certain
n-value (chosen to be 10 in practice). For recombination into levels
with n$>$10, the non-resonant contribution, relative to the resonant
contribution per unit energy is negligible owing to the density of
resonances as n $ \longrightarrow \infty $. In that range we employ a
precise theoretical treatment of DR based on multi-channel quantum defect
theory and the CC approximation (Bell \& Seaton 1985, Nahar \& Pradhan 
1994) to compute the recombination cross section.

 Among the problems that manfiest themselves in the CC
photoionization/recombination calculations are: the accuracy and
convergence of the
eigenfunction expansion for the ion, relativistic fine structure effects,
the contribution from non-resonant recombination into  high-n levels as $E
\longrightarrow 0; n \longrightarrow \infty$, and resolution of
narrow resonances with increasing n and/or $\ell$, and radiation damping
thereof.

 Experimental work is of importance in ascertaining the accuracy of
theoretical calculations and the magnitude of various associated
effects, since most of the photoionization/recombination data can 
only be calculated theoretically. In
recent years there has also been considerable advance in the
measurements of (e~+~ion) recombination cross sections on ion
storage rings (e.g. Kilgus et al, 1990, 1993, Wolf et al. 1991), 
and photoionization cross sections using accelerator based photon 
light sources (R. Phaneuf et al., private communication). We compare the CC
calculations for both atomic processes with the latest experimental
data.

\section{Theory}
 The CC approximation takes account of the important coupling between
the energtically accessible states of the ion in the (e~+~ion)
system. The target ion
is represented by an $N$-electron system, and the total wavefunction
expansion, $\Psi(E)$, of the ($N$+1) electron-ion system of symmetry $SL\pi$ or
$J\pi$ may be represented in terms of the target eigenfunctions as:

\begin{equation}
\Psi(E) = A \sum_{i} \chi_{i}\theta_{i} + \sum_{j} c_{j} \Phi_{j},
\end{equation}

\noindent
where $\chi_{i}$ is the target wavefunction in a specific state
$S_iL_i\pi_i$ or $J_i\pi_i$ and $\theta_{i}$ is the wavefunction for the
($N$+1)-th electron in a channel labeled as
$S_iL_i(J_i)\pi_ik_{i}^{2}\ell_i(SL\pi \ {\rm or}\ J\pi))$;
$k_{i}^{2}$ being its incident kinetic energy. $\Phi_j$'s are the correlation
functions of the ($N$+1)-electron system. Bound and continuum
wavefunctions for the (e~+~ion) system are obtained on solving the CC
equations at any total energy E $<$ 0 and E $>$ 0 respectively.
The coupling between the eigenfunctions of
the energetically inaccessible states of the ion ('closed channels'),
and the accessible states
('open channels'), gives rise to resonance phenomena, manifested as
infinite Rydberg series of resonances converging on to the excited
states of the target ion.

 With the bound and the continuum (free) states of the (e~+~ion) system,
atomic cross sections may be calculated for electron impact
excitation (EIE), photoionization, and recombination --- the free
and the bound-free processes. Radiative transition (bound-bound)
probabilities may also be obtained.

The R-matrix method, and its relativistic extension the Breit-Pauli
R-matrix method (BPRM), are the most efficient means of solving the CC
equations, enabling  in particular the resolution of the resonances in
the cross sections at a large number of energies.

\section{Photoionization}

 Photoionization calculations are carried out for all levels ($n,\ell,
SL\pi,SLJ\pi$, with $n \leq 10, \ell = n-1$. Typically this means
several hundred bound levels of each atom or ion. The cross section for
each level is delineated at about thousand energies, or more, to map out
the resonance structure in detail. In recent studies, photoionization
cross sections of all ions of C, N, and O  were computed (Nahar and Pradhan
1997, Nahar 1998) with more extended eigenfunction expansions, resolution
of resonances, and number of levels, than the earlier OP data.
Overall, new photoionization data for over 40 atoms and ions,
with improvements over the OP data (e.g. currently in TOPbase,
Cunto et al. 1993), has now been calculated for: low
ionization stages of iron: Fe~I, II, III, IV, and V (references in
Bautista and Pradhan 1998), Ni~II (Bautista 1999), the C-sequence
ions (Nahar \& Pradhan 1991,1992), the Si-sequence ions (Nahar \& Pradhan 
1993). Unified (e~+~ion) photo-recombination cross sections
and rates (total and level-specific) have also been obtained,
as discussed below.

\subsection{Comparison with experiments}

The recent ion-photon merged beam experiment by Kjeldsen et al. (1999) 
on the photoionization cross sections of the ground state of C II shows 
an extremely rich and detailed resonance structures (Fig. 1). There is 
excellent agreement between the theory and experiment, both in terms of 
magnitude and details of the background and resonances. However, the 
theoretical calculations were in LS coupling, neglecting fine structure, 
that clearly manifests itself in the additional peaks seen in the
experimental cross sections (new relativistic calculations are in
progress).

\begin{figure}
\vspace*{-2.0cm}
\centering
\psfig{figure=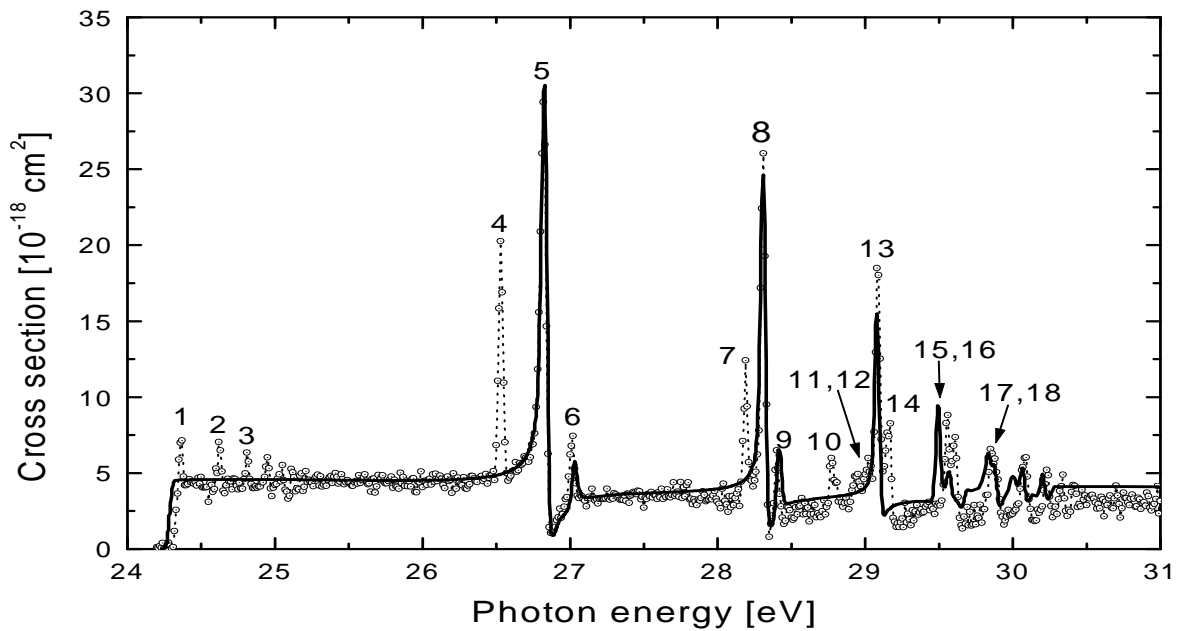,height=10.0cm,width=18.0cm}
\vspace*{-1cm}
\caption{Photoionization cross sections of the ground state of C II:
experiment (open circle, Kjeldsen et al. 1999), theory (solid curve,
Nahar 1995)}
\end{figure}

New photoionization experiments have been carried out for
positive atomic ions at the Advance Light Source (ALS) in Berkeley, where
a photon light source is used on an accelerator that produces the ion beams. 
These extremely high resolution measurements provide an unprecedented 
check on the details of the theoretical cross sections, particularly 
resonance structures and fine structure effects.
 Fig. 2 compares the O~II cross section from theory (Nahar 1998), and
experiments at the ALS done by the Reno group
headed by R. Phaneuf. The experimental cross sections include not only
the photoionization of the ground state $2s^22p^3 \ \ (^4S^o)$ but also
the metastable excited states $2s^22p^3 \ \ (^2D^o, ^2P^o)$.

$$ \\ h\nu + O~II (2s^22p^3 \ \ ^4S^o,^2D^o,^2P^o) \longrightarrow e + O~III
(2s^22p^2 \ \ ^3P,^1D,^1S) \\ $$

 The complicated features arise from several series of resonances
converging on to the excited states of the residual ion O~III. There is
very good agreement between the CC calculations and experiment,
verifying the often expressed, but not heretofore established,
claim of about 10\% accuracy of the theoretical cross sections.
The situation may be more complicated, and the uncertainties larger, for
more complex atomic systems.

 {\it These results also show that metastable states may need to be included
in atomic photoionization models of astrophysical sources}.


\begin{figure}
\centering
\psfig{file=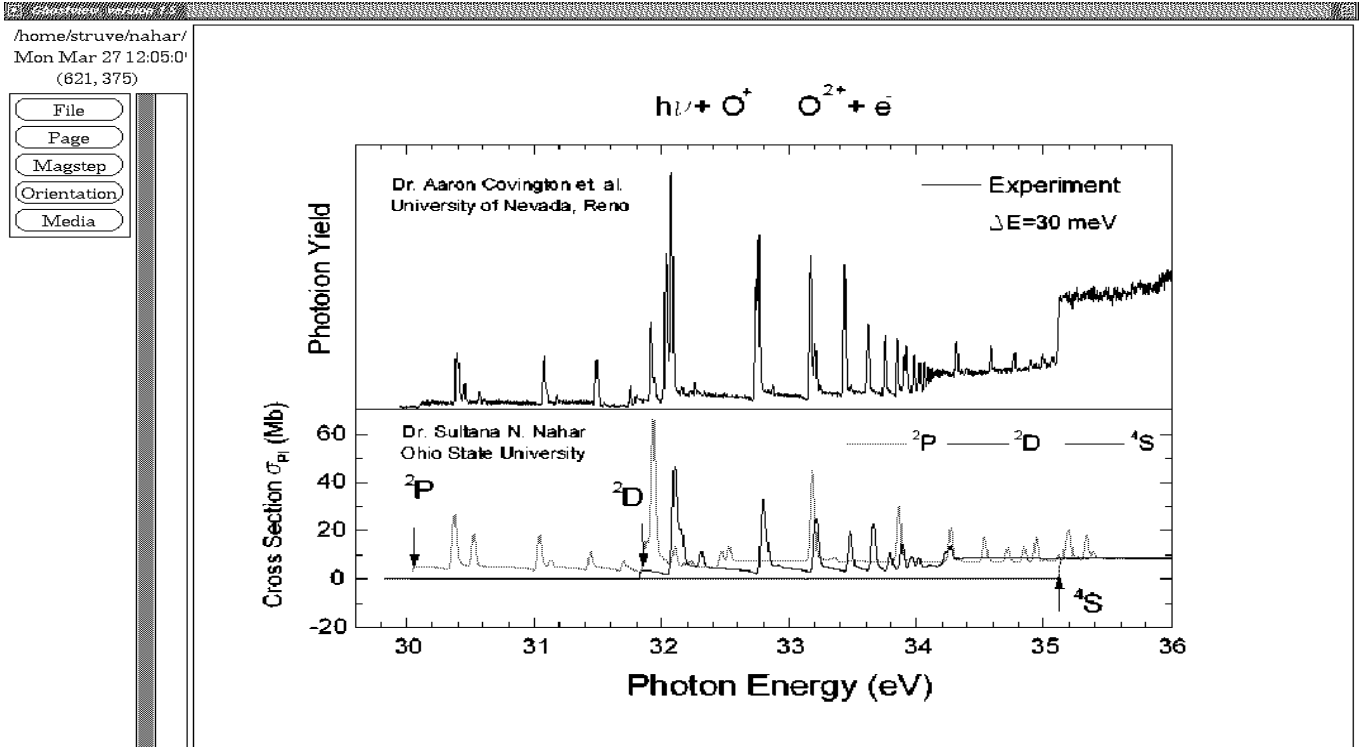,height=10.0cm,width=18.0cm}
\vspace*{-1.0cm}
\caption{Photoionization cross sections of the ground configurations of
O II: experiment (upper panel, Phaneuf et al.), theory (lower panel,
Nahar 1998).}
\end{figure}

\section{Unified method for (e + ion) recombination}

Photoionization calculations described above are for total
photoionization from a given level into all excited levels of the
residual ion. However, for photo-recombination the calculations must 
be repeated to obtain the cross section into the ground state of the 
ion alone. Detailed balance then applies precisely (Milne relation) as

\begin{equation}
\sigma_{\rm RC}(\epsilon) =
{\alpha^2 \over 4} {g_i\over g_j}{(\epsilon + I)^2\over
\epsilon}\sigma_{\rm PI},
\end{equation}
where $\sigma_{\rm RC} (i_o)$ is the photo-recombination cross section,
$\sigma_{\rm PI}$ is the photoionization cross section into the 
ground state $i_o$, $\alpha$ is the fine structure constant, $\epsilon$ 
is the photoelectron energy, and $I$ is the ionization potential in 
Rydberg atomic units. Recombination can take place into the ground or 
any of the excited recombined (e+ion) states. The contributions of 
these bound states to the total $\sigma_{\rm RC}$ are obtained by 
summing over the contributions from individual cross sections. 
$\sigma_{\rm RC}$ thus obtained from $\sigma_{\rm PI}$, including the 
autoionizing resonances, corresponds to the total (DR+RR) unified 
recombination cross section in an ab initio manner.

Recombination into the high-$n$ states, i.e.
$n_{\rm max} < n \leq \infty$, is computed assuming DR to dominate over
the non-resonant background contribution. The CC approximation can then
be used to calculate DR collision strengths $\Omega_{\rm
DR}$, as an extension of the theory of DR by Bell and Seaton (1985).
These two main parts of the unified recombination calculations, and
other parts, are described in detail in Zhang et al (1999).
For very highly charged ions, such as the H- and He-like ions with large
radiative decay rates for core transitions, radiation damping effects
can be significant. As in other CC calculations for excitation and
photoionization, resonances are resolved at a suitably fine mesh to enable
perturbative radiative damping, and to ensure that the neglected
resonances do not significantly affect the computed rates.
Relativistic fine structure is considered in the BPRM calculations for
highly charged ions.


\subsection{Comparison with experiments}

Although experimental results are available for relatively few ions in
limited energy ranges, and mostly for simple atomic systems such 
as the H-like and He-like ions, they are useful for the 
calibration of theoretical cross sections. Zhang et al (1999) have 
compared in detail the BPRM cross sections with experimental data 
from ion storage rings for $e~+~C~V \longrightarrow C~IV, e~+~C~VI 
\longrightarrow C~V, e~+~OVIII \longrightarrow O~VII$, with close 
agreement in the entire range of measurements for both the 
background (non-resonant) cross sections and resonances. The reported 
experimental data is primarily in the region of low-energy resonances 
that dominate recombination (mainly DR) with H- and He-like ions. 
The recombination rate coefficients, $\alpha_R$, obtained 
using the cross sections calculated by Zhang et al. agree closely with 
those of Savin (1999) who used the experimental cross sections to obtain 
'experimentally derived DR rates'. However, these rates do not include 
contributions from  much of the low energy non-resonant RR and very high
energy regions. The total unified $\alpha_R(T)$ which include all possible
contributions is, therefore, somewhat higher than that obtained from 
limited energy range. In Fig. 3, the solid curve corresponds to the total
unified $\alpha_R$. The dotted curve and the dot-long-dash
curves are the
rates using cross sections from Zhang et al (1999) and Savin (1999)
respectively, in the limited energy range in experiments
(the two curves almost merge).
The short-and-long dash curve is the total $\alpha_R$ in LS 
coupling (Nahar \& Pradhan 1997) which, at high temperatues, is higher
than the new BPRM rates including fine structure and radiation damping (solid
curve).
The dot-dash curve is the DR rate by Badnell et al (1990), which is lower than 
the others. the dashed and the long-dashed curves are RR rates by
Aldrovandi \& Pequignot (1973), and Verner and Ferland (1996); the 
latter agrees with the present rates at lower temperatures.

\begin{figure}
\vspace*{-2cm}
\centering
\psfig{figure=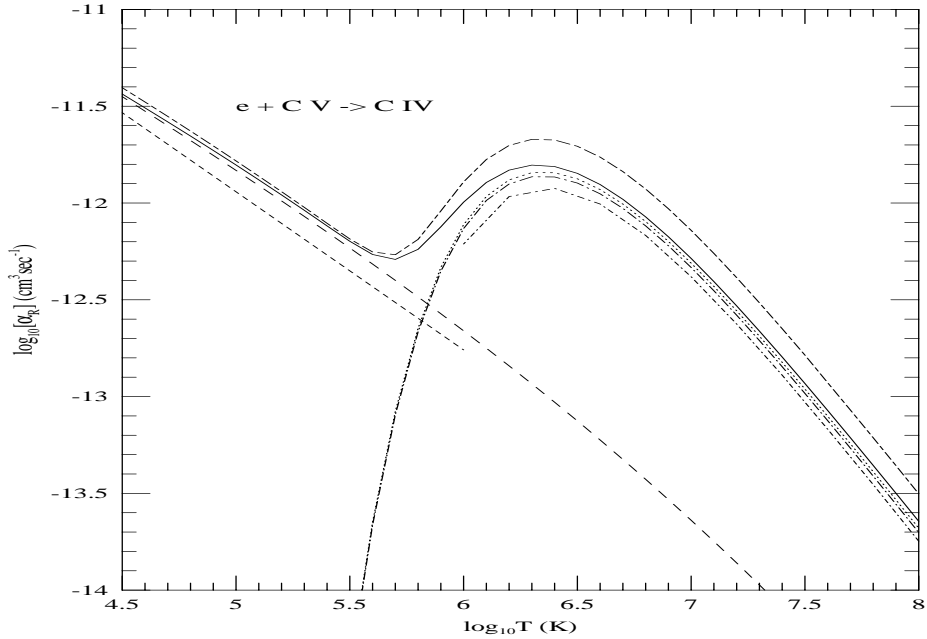,height=10.0cm,width=15.0cm}
\vspace*{-1cm}
\caption{e + C V $\longrightarrow$ CIV. Total unified rate coefficients:
BPRM with fine structure -- solid curve, LS coupling -- short and long 
dashed curve; using cross sections from Zhang et al (1999) -- dotted; 
Savin (1999) -- dot-long dash curve; DR rates by Badnell et al (1990) 
-- dot dash curve; RR rates: Aldrovandi and Pequignot (1973) -- 
short-dash; Verner and Ferland (1996) -- long-dash}
\end{figure}

\subsection{Photoionization/recombination of Fe~XXV}

Fe~XXV is one of the most important ions in X-ray spectroscopy (see the
review article by Pradhan in this volume). We 
have completed the calculations for: (i) photoionization cross
sections for fine structure levels up to n = 10, including those for 
ionization into the ground level, and (ii)
total and level-specific unified recombination cross sections and
rate coefficients.

The most commonly observed lines of Fe XXV correspond to the K-$\alpha$ 
transitions between the $n$= 1 and 2 levels: (1) the 'z' line
$1s2s(^3S_1)-1s^2(^1S_0)$, (2) the 'y' line, $1s2p(^3P^o_1)-1s^2(^1S_0)$,
(3) the 'x' line,  $1s2p(^3P^o_2)-1s^2(^1S_0)$, and (4) the 'w' line
$1s2p(^1P^o_1)-1s^2(^1S_0)$. Recombination rate coefficients into 
these levels are given in 
Fig. 4; these vary smoothly with temperature, 
except a slight "shoulder" at high temperature due to DR.

\begin{figure}
\vspace*{-2cm}
\psfig{figure=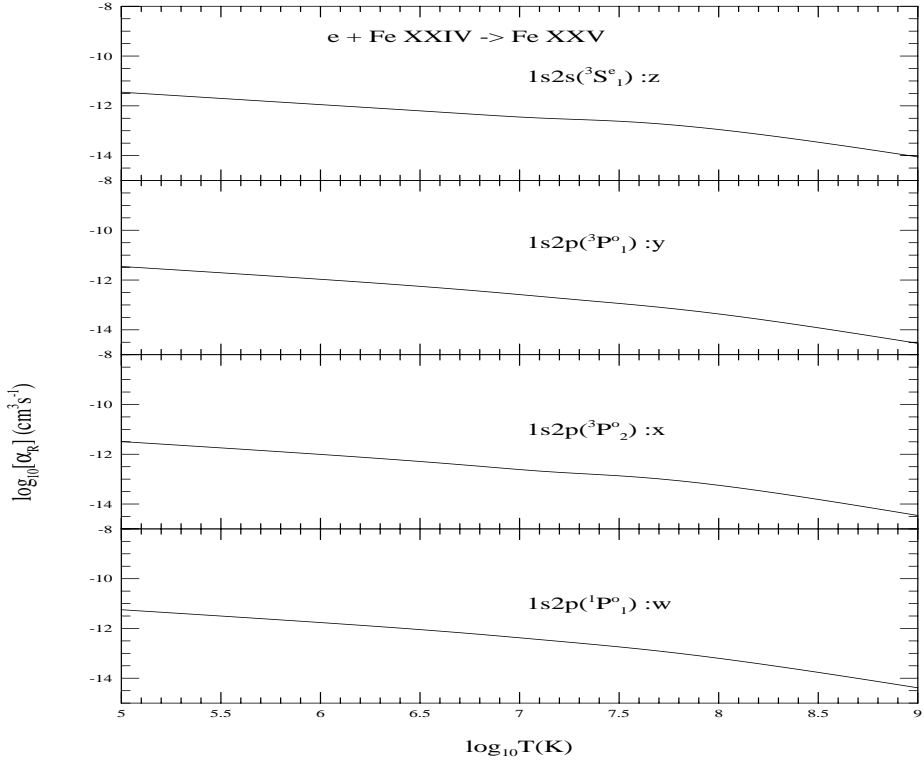,height=12.0cm,width=15.0cm}
\vspace*{-1cm}
\caption{Level specific recombination rate coefficients for the 
K-$\alpha$ lines of Fe XXV.} 
\end{figure}

\section{Ionization equilibrium}

 The new photoionization/recombination data  should enable more accurate
calculations for {\bf photoionization equilibrium}

\begin{equation}
  \int_{\nu_0}^{\infty} \frac{4 \pi J_{\nu}}{h\nu} N(X^{z})
\sigma_{PI}(\nu,X^{z}) d\nu = \sum_j N_e N(X^{z+1}) \alpha_R(X_j^{z};T),
\end{equation}

and, {\bf coronal equilibrium}

\begin{equation}
C_I(T,X^{z}) N_e N(X^{z}) = \sum_j N_e N(X^{z+1}) \alpha_R(X_j^{z};T),
\end{equation}

\noindent
where $\alpha_R(X_j^{z};T)$ is the total electron-ion recombination rate
coefficient of the recombined ion of charge $z$, $X_j^{z}$, to state j
at electron temperature T, $C_I$ is the rate coefficient for electron
impact ionization, and $\sigma_{PI}$ is the photoionization cross
section
evaluated at photon frequency $\nu$ and convoluted with the isotropic
radiation density J$_{\nu}$ of the source; N$_e$, $N(X^{z+1})$, and
$N(X^{z})$ are the densities for the free electrons, and the recombining
and recombined ions respectively.

Coronal ionization fractions for C,N,O using the unified recombination
rates are also computed (Nahar and Pradhan 1997; Nahar 1999).

\section{Conclusion}

We carry out ab initio large scale close coupling R-matrix
calculations for (i) photoionization cross sections, and (ii) 
electron-ion recombination rate coefficients. The predicted 
theoretical features in $\sigma_{PI}$ are being observed in the recent 
sophisticated experiments.

The unified method for (e+ion) recombination has been benchmarked with 
available experimental measurements. Our study of unified electron-ion 
recombination rates exhibit a general pattern with temperature.
Although generally applicable to all systems, the close coupling
unified method is especially suitable for the strong
coupling cases where the broad and overlapping resonances dominate
the near-threshold region in the electron-ion recombination process,
and other methods may not be accurate.

Total and state-specific unified recombination rate coefficients and
photoionization cross sections are available for about 40 atoms and 
ions: \\
Carbon: C I, C II, C III, C IV, C V, C VI \\
Nitrogen: N I, N II, N II, N IV, N V, N VI, N VI \\
Oxygen: O I, O II, O III, O IV, O V, O VI, O VII, O VII \\
C-like: F IV, Ne V, Na VI, Mg VII, Al VIII, Si IX, S XI \\
Si and S: Si I, Si II, S II, S III, Ar V, Ca VII \\
Iron: Fe I, Fe II, Fe III, Fe IV, Fe V, Fe XIII, Fe XXV 

 These photoionization/recombination datasets are self-consistent,
and should yield more accurate astrophysical photoionization models.

\vskip 0.2in
\begin{center}
{\bf 4. Acknowledgements}
\end{center}

This work is supported partially by the NSF and NASA.

\vskip 0.2in
\begin{center}
{\bf REFERENCES}
\end{center}

\noindent
Aldrovandi, S.M.V. \& Pequignot, D. 1973, Astron. Astrophys. 25,
137 \\
Badnell, N.R., Pindzola, M.S., \& Griffin, D.C. 1990, Phys.
Rev. A 41, 2422 \\
Bautista, M.A. 1999, Astron. Astrophys. Suppl. 137, 529 \\
Bautista, M.A. \& Pradhan A.K. 1988, Astrophys. J. 492, 650 \\
Bell, R.H.  and Seaton, M.J. 1985, J. Phys.  B, 18, 1589 \\
Berrington K.A., Burke P.G., Butler K., Seaton M.J., Storey
P.J., Taylor K.T., Yu Yan, 1987, J.Phys.B 20, 6379 \\
Burgess, A. 1964, Astrophys. J. 141, 1588 \\
Burke P.G. \& Seaton M.J. 1984, J. Phys. B 17, L683 \\
Cunto W C, Mendoza C, Ochsenbein F and Zeippen C J, 1993. A\&A,
275, L5\\
Hummer D.G., Berrington K.A., Eissner W., Pradhan A.K, Saraph H.E.,
Tully J.A., 1993, A\&A, 279, 298 \\
Kilgus G, Berger J, Blatt P, Grieser M, Habs D, Hochadel B,
Jaeschke E, Kr\"{a}mer D, Neumann R, Neureither G, Ott W, Schwalm D, 
Steck M, Stokstad R, Szmola R, Wolf A, Schuch R, M\"{u}ller A and 
W\"{a}gner M 1990 Phys. Rev. Lett. 64, 737 \\
Kilgus G, Habs D, Schwalm D, Wolf A, Schuch R and Badnell N R
1993 Phys. Rev. A 47, 4859 \\
Kjeldsen H., Folkmann F., Hensen J.E., Knudsen H., Rasmussen M.S., 
West J.B., Andersen T., 1999, Astrophys. J. 524, L143 \\
Luo D., Pradhan A.k., Saraph H.E., Storey P.J., Yan Y. 1989, J. Phys. B
22, 389 \\
Nahar, S.N. 1998, Phys. Rev. A 58, 4593 \\
Nahar, S.N. 1999, ApJS 120, 131 \\
Nahar, S.N., \& Pradhan, A.K. 1991, Phys. Rev. A 44, 2935 \\
Nahar, S.N., \& Pradhan, A.K. 1992, Phys. Rev. A 45, 7887 \\
Nahar, S.N., \& Pradhan, A.K. 1993, J. Phys. B 26, 1109 \\
Nahar, S.N., \& Pradhan, A.K. 1994, Phys. Rev. A 49, 1816 \\
Nahar, S.N., \& Pradhan, A.K. 1997, ApJS 111, 339 \\
Savin, D.W. 1999, Astrophys. J. 523, 855
{\it The Opacity Project 1 \& 2}, compiled by the Opacity Project team 
(Institute of Physices, London, UK, 1995,1996) \\
Wolf A, Berger J, Bock M, Habs D, Hochadel B, Kilgus G,
Neureither G, Schramm U, Schwalm D, Szmola E, M\"{u}ller A, Waner M,
and Schuch R 1991 Z. Phys. D Suppl. 21, 569 \\
Verner, D.A. \& Ferland G. 1996, Astrophys. J. Suppl. 103, 467 \\
Zhang, H.L., Nahar, S.N., \& Pradhan, A.K. 1999, J. Phys. B 32, 1459 \\

\end{document}